\documentclass[aps,pra,twocolumn,superscriptaddress,showpacs,amsmath,amssymb,longbibliography,10pt,floatfix]{revtex4-1}
\usepackage{graphicx}
\usepackage[breaklinks,colorlinks,linkcolor=blue,citecolor=blue,urlcolor=blue]{hyperref}

\usepackage[utf8]{inputenc}
\usepackage[english]{babel}
\usepackage{amsmath}
\usepackage{xcolor}

\definecolor{nred} {RGB}{224,0,0}
\definecolor{nblue} {RGB}{28,130,185}

\usepackage{ulem} % sout

\begin{document}

\title{Vanishing Wilson ratio as the hallmark of quantum spin-liquid models} 
\author{P. Prelov\v{s}ek}
\affiliation{Jo\v zef Stefan Institute, SI-1000 Ljubljana, Slovenia}
\affiliation{Faculty of Mathematics and Physics, University of Ljubljana, 
SI-1000 Ljubljana, Slovenia}
\author{K. Morita}
\author{T. Tohyama}
\affiliation {Department of Applied Physics, Tokyo University of Science, 
Tokyo 125-8585, Japan}
\author{J. Herbrych}
\affiliation{Department of Theoretical Physics, Faculty of Fundamental Problems 
of Technology, Wroclaw University of Science and Technology, 50-370 Wroclaw, Poland}

\begin{abstract}
We present numerical results for finite-temperature $T>0$ thermodynamic 
quantities, entropy $s(T)$, uniform susceptibility $\chi_0(T)$ and the Wilson 
ratio $R(T)$, for several isotropic $S=1/2$ extended Heisenberg models which are 
prototype models for planar quantum spin liquids. We consider in this context 
the frustrated $J_1$-$J_2$ model on kagome, triangular, and square lattice, as 
well as the Heisenberg model on triangular lattice with the ring exchange. Our 
analysis reveals that typically in the spin-liquid parameter regimes the 
low-temperature $s(T)$ remains considerable, while $\chi_0(T)$ is reduced 
consistent mostly with a triplet gap. This leads to vanishing $R(T \to 0)$, 
being the indication of macroscopic number of singlets lying below triplet 
excitations. This is in contrast to $J_1$-$J_2$ Heisenberg chain, where 
$R(T \to 0)$ either remains finite in the gapless regime, or the singlet and 
triplet gap are equal in the dimerized regime.
\end{abstract}
%================================================================================
\maketitle

%================================================================================
\section {Introduction} 

Various frustrated $S=1/2$ Heisenberg models (HM) have been subject of intensive 
theoretical studies in last decades in connection with the possibility of 
spin-liquid (SL) ground state (g.s.). These efforts have been recently 
strengthened by the discovery of several classes of insulating materials 
revealing low-energy spin excitations behaving as quantum SL without any 
magnetic order down to low temperatures (for reviews see 
\cite{lee08,balents10,savary17}). Among isotropic $S=1/2$ two-dimensional (2D) 
models most numerical evidence for the SL g.s. accumulated for the 
antiferromagnetic (AFM) HM on the kagome lattice (KL) 
\cite{mila98,budnik04,singh07,yan08,lauchli11,iqbal11,depenbrock12}, but as well 
for $J_1$-$J_2$ HM on the square lattice (SQL) 
\cite{capriotti00,mambrini06,jiang12,gong14,morita15,morita16,wang18,liu18}, 
$J_1$-$J_2$ HM on the triangular lattice (TL) 
\cite{kaneko14,zhu15,hu15,iqbal16,wietek17,prelovsek18} and the HM on TL with 
ring exchange \cite{misguich99,motrunich05}. While the character of the g.s. and 
its properties still offer several controversies and challenges, it is even less 
known about finite-temperature $T>0$ behavior of several basic quantities. At 
least some of them have been already measured in experiments on SL materials and 
can thus serve as a test whether and to what extent actual materials can be 
accounted for by theoretical models. 

Among measurable spin properties are thermodynamic quantities as the uniform 
magnetic susceptibility $\chi_0(T)$, magnetic (contribution to) specific heat 
$C_V(T)$ and related spin entropy density $s(T)$. They are crucial to pinpoint 
the different characters and scenarios of SL behavior, in particular whether 
materials follow gapped or gapless SL. These quantities are mostly extracted 
from experiments on KL systems, the prominent example being herbertsmithite 
\cite{mendels07,olariu08,han12,fu15,norman16}, but also related compounds in the 
same class \cite{hiroi01,fak12,li14,gomilsek16,feng17,zorko19}. Another example 
are organic compounds where the relevant lattice is triangular 
\cite{shimizu03,shimizu06,itou10,zhou17} as well as charge-density-wave system 
1T-TaS$_2$, recently established as SL with composite $S=1/2$ spins on TL 
\cite{klanjsek17,kratochvilova17,law17,he18}. The basic spin exchange scale in 
most of these systems is modest and, as a consequence, the whole $T$ range is 
experimentally accessible which allows for the test of the whole range of spin 
excitations. Nevertheless, it should be noted that properties at lowest $T$ might be 
influenced by additional mechanisms such as Dzyaloshinski-Moriya interaction 
\cite{rigol07,cepas08,zorko08}, interlayer couplings and random effects 
\cite{kawamura19}.

It has been rather well established with elaborate exact-diagonalization (ED) 
and series-expansion studies of the HM with nearest-neighbor (n.n.) exchange on 
KL \cite{mila98,budnik04,singh07,singh08,lauchli11,lauchli19}, that lowest 
excitations are singlets dominating over the triplet excitations, for which most 
ED studies reveal a finite spin (triplet) gap $\Delta_t >0$ although there are 
numerical indications also for gapless scenario \cite{iqbal13,he17}. It has been 
recently shown \cite{prelovsek19} that the same scenario can be traced via the 
temperature-dependent Wilson ratio $R(T)$ in $J_1$-$J_2$ HM on TL including 
the next-nearest-neighbor (n.n.n.) exchange $J_2>0$ in the regime where the SL 
g.s. is expected \cite{kaneko14,zhu15,hu15,iqbal16}. This is in contrast with 
the triplet ($S=1$) magnon excitation being the lowest excitations in an ordered 
AFM. It is also qualitatively different from the scenario for the basic 
one-dimensional (1D) HM with gapless spinon excitations. 

In the following we present numerical results for $s(T)$, $\chi_0(T)$ and 
$R(T)$, which reveal that the vanishing $R(T \to 0)$ is quite generic 
property of a wide class of isotropic 2D Heisenberg models in their range of 
(presumable) SL parameter regimes. In this context we generalize previous 
numerical $T>0$ studies of HM on KL \cite{misguich07,schnack18} to include also 
the n.n.n. exchange $J_2 \neq 0$ and upgrade results for the $J_1$-$J_2$ HM on 
TL \cite{prelovsek18}, now studying also the HM on TL with the ring exchange, as 
well as another standard model of SL, i.e., frustrated $J_1$-$J_2$ HM on SQL. 
Results in the SL regimes confirm singlets as dominating low-energy 
excitations. For comparison we present results also for 1D $J_1$-$J_2$ 
Heisenberg chain, which serve as the reference, depending on $J_2/J_1$, either 
for the gapless spinon Fermi-surface (SFS) and valence-bond (VB) solid 
scenarios. Still, we  show that results appear (as expected) qualitatively different 
from considered 2D models.

Investigated models have their particular features and challenges, nevertheless our 
results on thermodynamic properties reveal quite universal properties in their 
(presumable) SL regimes which put also restrictions on the SL scenarios 
explaining their low-$T$ behavior. In particular, very attractive scenario of 
gapless SL with SFS excitations requires finite g.s. Wilson ratio 
$R_0= R(T \to 0) >0$. The latter is realized in 1D HM, but does not appear to be 
the case in planar models. Observed enhanced low-$T$ entropy $s(T)$ and related 
vanishing $R_0=0$ demonstrate the dominant role of singlet excitations over the 
triplet ones \cite{waldtmann98,singh08}, but still offer several possibilities. 
While it is hard to exclude the scenario of VB solid (crystal) with broken 
translational symmetry \cite{singh07,singh08}, it is more likely that the g.s. 
in the SL regime does not break the translation symmetry and all correlations 
are short-ranged, i.e. revealing a scenario of VB (or dimer) liquid. On the 
other hand, it is well possible that considered models might not be enough to 
represent the SL real materials, in particular not in their low-$T$ regime.

The paper is organized as follows: In Sec.~II we introduce $T$-dependent Wilson 
ratio $R(T)$ and comment different scenarios for its low-$T$ behavior. 
In Sec.~III we present numerical methods used to evaluate thermodynamic 
quantities, but also lowest spin excitations in 1D and 2D models. As the 
test of methods as well as of concepts we present in Sec.~IV results for 1D 
$J_1$-$J_2$ Heisenberg chain. The central results for various 2D frustrated HM 
models are presented and analyzed in Sec.~V, and summarized in Sec.~VI.

%================================================================================
\section{Temperature-dependent Wilson ratio}

Besides thermodynamic quantities: uniform magnetic susceptibility $\chi_0(T)$ 
and the entropy density $s(T)$, together with related specific heat 
$C_V(T)=T ds/dT$, it is informative to extract also their quotient in the form 
of temperature-dependent Wilson ratio $R(T)$, defined as \cite{jaklic00,prelovsek19},
\begin{equation}
R(T)= \frac{4 \pi^2 T \chi_0(T) }{3 s(T)}\,,
\label{rw}
\end{equation}
being dimensionless quantity assuming (theoretical) units $k_B= g \mu_B =1$. It should be 
reminded that the standard quantity is the (zero-temperature) Wilson ratio as 
$R_W = 4 \pi^2 \chi^0_0 / (3 \gamma) $ where $\chi^0_0=\chi_0(T=0)$ and 
$\gamma = \mathrm{lim}_{T \to 0} [C_V/T]$. $R_W$ has its usual application and 
meaning in the theory of Fermi liquids and metals, as well as in gapless spin 
systems \cite{ninios12}. We note that in normal Fermi-liquid-like systems where 
$s= C_V=\gamma T$ the definition, Eq.~(\ref{rw}), coincides at $T \to 0$ with 
the standard $R_W$. Although at low $T$ (in most interesting cases) both $s(T)$ 
and $C_V(T)$ have the same functional $T$ dependence, it is more convenient to 
employ in Eq.~(\ref{rw}) the entropy density $s(T)$ being monotonously 
increasing function. 

It should be also pointed out that $R(T)$ is a direct measure of the ratio of 
the density of excitations with finite $z$ component of total spin 
$S_{tot}^z \neq 0$ relative to density of all (spin) excitations, including 
$S_{tot}^z =0$, as measured by $s(T)$. To make this point evident we note that 
$\chi_0(T) = \langle (S^{tot}_z)^2 \rangle /(NT)$ where $N$ is the number of 
lattice sites, so that
\begin{equation}
R= \frac{4 \pi^2 \langle (S_{tot}^z)^2\rangle }{3 N s}\,. \label{r1}
\end{equation}
From above expression it is also follows that $R(T)$ has a well defined 
high-$T$ limit which is for isotropic $S=1/2$ HM 
$R(T \to \infty) = \pi^2/(3 \ln2) = 4.746$. 

Moreover, $R_0 \equiv R(T \to 0)$ can differentiate between distinct scenarios: 

\noindent a) In the case of magnetic long-range order (LRO), e.g., for AFM in HM 
on SQL, at $T \to 0$ one expects in 2D isotropic HM 
$\chi_0(T\to 0) = \chi_0^0 >0$ (where the finite value can be interpreted as the 
contribution of spin fluctuations transverse to the g.s. magnetic order) whereas 
effective magnon excitations lead to $s \propto T^2$~\cite{manousakis91}, so 
that $R_0 \propto 1/T \to \infty$, 

\noindent b) In a gapless SL with large SFS one would expect Fermi-liquid-like 
finite $R_0 \sim 1$ \cite{balents10,zhou17,law17}. The evident case for such 
scenario, as for reference considered later on, is the simple Heisenberg chain 
where $R_0 = 2$~\cite{johnston00}, in contrast to the value $R_0=1$ for 
noninteracting Fermi systems. 

\noindent c) Vanishing $R_0 \to 0$, or more restricted from Eq.~(\ref{rw}) 
$R_0 \propto T^\eta$ with $\eta \geq 1$, would indicate that low-energy singlet 
excitation dominate over the triplet ones \cite{singh08,balents10,lauchli19}. In 
the following we find numerical evidence that this appears to be the case in the 
SL parameter regime of considered 2D frustrated isotropic HM.

Within the last scenario one should still differentiate different possibilities 
with respect to gapless spin systems or systems with the gap. One option for SL 
is that both singlet and triplet excitations are gapped, but the effective 
triplet gap is larger $\Delta_t>\Delta_s$ (in the limit of large systems 
$N \to \infty$) which would lead (in a simplest approximation) to 
$R_0 \propto T^\eta \exp[-(\Delta_t- \Delta_s)/T] \to 0$. More delicate case 
could be when $\Delta_t = \Delta_s = \Delta$. Then Eq.~(\ref{rw}) offers several 
scenarios with, e.g., $R(T<\Delta) \propto T^\eta$. Such situation appears, 
e.g., for 1D chain $J_1$-$J_2$ model around the Mazumdar-Ghosh point 
$J_2/J_1=0.5$. Since $s(T)$ measures both singlet and triplet excitations (as 
well as higher $S_{tot} >1$) possible case $\Delta_s > \Delta_t$ should be 
similar to the previous scenario.

When classifying options for $T \to 0$ we should also consider the possibility 
of VB solid (crystal), i.e., the g.s. with broken translational symmetry. In 
finite systems (with short-range spin correlations) the signature of VB solid 
should be the degenerate or (due to finite-size effects) nearly degenerate g.s. 
with degeneracy $N_d > 1$. This should be reflected in a finite g.s. entropy for 
finite system with $N$ sites,
\begin{equation}
s_0 \equiv s(T \to 0) = \frac{1}{N} \ln N_d. \label{s0}
\end{equation} 
Such remnant $s_0 >0$ does not contribute to $C_V(T)$ and moreover vanishes in 
the limit $N \to \infty$. A clear VB solid case is 1D $J_1$-$J_2$ HM in the 
dimerized regime where $N_d=2$. It then makes sense to consider in the 
evaluation of $R(T)$, Eq.~(\ref{rw}), besides full $s(T)$ also reduced one 
$\tilde s=s-s_0$. Still, it is not always straightforward to fix proper $N_d$ in 
finite-size systems.

%================================================================================
\section{Methods}

We calculate entropy density $s(T)$, uniform susceptibility $\chi_0(T)$ and via 
Eq.~(\ref{rw}) the Wilson ratio $R(T)$, using the finite-temperature Lanczos 
method (FTLM) \cite{jaklic94,jaklic00}, previously used in numerous studies of 
$T>0$ static and dynamical properties in various models of correlated electrons 
\cite{prelovsek13}. Since in the case of considered thermodynamic quantities 
only conserved quantities are involved, in particular the Hamiltonian $H$ and 
$S_{tot}^z$, the memory and CPU time requirement for given system size $N$ are 
essentially that of the Lanczos procedure for the g.s., provided that we scan 
over all (different) symmetry sectors $S_{tot}^z$ and wave-vector ${\bf q}$ due 
to translational symmetry and periodic-boundary conditions (p.b.c.), in case of 
the code with translational symmetry. A modest additional sampling $N_s$ over 
initial wave-functions is then used. Limitations of the present method are 
given by the size of the many-body Hilbert space with $N_{st}$ basis states 
which can be handled efficiently within the FTLM, restricting in our study 
lattice sizes to $N \leq 36$. In the following we use two FTLM codes for 
considered models:

\noindent a) To calculate largest systems with $N = 36$ sites for the 2D TL, KL 
as well as SQL $J_1$-$J_2$ HM with $N_{st} \sim 10^{10}$, we develop a code that 
equips a technique to save the memory for the Hamiltonian by dividing $H$ 
into two subsystems. In addition, to improve the accuracy, we use replaced FTLM 
technique \cite{morita19}.

\noindent b) The code for more modest computers takes into account translational 
symmetry, thus able to reach $N_{st} < 10^7$ and sizes $N \leq 30$, was used for 
the 1D HM chain and the TL with ring exchange.

When discussing the accuracy of FTLM results we have to distinguish results for 
given system from finite-size effects due to restricted $N$. The central 
quantity evaluated is the grand-canonical sum \cite{jaklic94,jaklic00},
\begin{equation}
Z(T)=\mathrm{Tr }~\mathrm{exp}[-(H-E_0)/T],
\end{equation}
where $E_0$ is the g.s. energy. For reachable systems FTLM provides accurate 
results provided that we use modest random sampling $N_s \leq 30$ over (random) 
initial wave-functions. This is in particular important to get correct low-$T$ 
limit, i.e., $Z(T=0) =1$ in the case of non-degenerate g.s. \cite{morita19}. 
The main restriction  of FTLM results are, however, reachable $N$ and related finite-size effects most 
pronounced at $T \to 0$:

\noindent a) In isotropic HM with $T \to 0$ LRO (in dimension $D\geq 2$), or 
long-range spin correlations in 1D, spin excitations are gapless in the 
thermodynamic limit. Such case is correlated with finite-size effects in 
evaluated quantities. One can expect that results reach the $N \to \infty$ 
validity only for $Z>Z^*=Z(T_{fs}) \gg 1$. Since $Z$ is intimately related to 
entropy,
\begin{equation}
s = \frac{1}{N} \left( \ln Z + \frac{\langle H \rangle - E_0 }{T} \right)\,,
\label{s}
\end{equation}
the criterion for $T_{fs}$ can be the smallest value for $s$. Actually, in 
reached systems $N \sim 36$ we get estimate $s(T_{fs}) \sim 0.07 - 0.1$ (see, 
e.g., the finite-size analysis in \cite{schnack18}). In such systems $s(T)$ and 
$\chi_0(T)$ results at $T<T_{fs}$ are dominated by finite-size effects and are 
not representative for $N \to \infty$. In any case, due to frustration and 
consequently enhanced $s(T \ll J_1)$ in SL models FTLM generally allows to 
reach lower $T_{fs}$. E.g., while for HM on an unfrustrated SQL (even 
at largest $N=36$) $T_{fs} \sim 0.4 J_1 $ \cite{jaklic00,prelovsek13}, SL models 
allow for considerably lower $T_{fs} \leq 0.1 $ \cite{schnack18,prelovsek18}.

\noindent b) For systems with only short-range spin correlations one can reach 
situation where spin correlation length (even at $T \to 0$) is shorter that the 
system length, $\xi \le L$. In such a case, FTLM has no obvious restrictions 
even at $T \to 0$, so $T_{fs} \sim 0$. This can be the situation for gapped SL, 
including some examples discussed further on. 

Besides thermodynamic quantities, it is also instructive to monitor directly 
lowest excited states and their character. For the largest 2D $N=36$ systems 
excited states are obtained within ED (without translational symmetry) by 
eliminating Lanczos-ghost states while comparing results for different number of 
Lanczos steps. For TL with ring exchange we employ ED results of systems with 
$N =28$ and evaluate the lowest (singlet and triplet) energies in different 
${\bf q}$ sectors.

In 1D models we use also density matrix renormalization group (DMRG) method to 
investigate the $J_1$-$J_2$ HM with open boundary conditions (o.b.c.). The 
method allows for accurate computation of the $S^z_{tot}=0$ g.s., and in the 
same way also first excited triplet state with $S^z_{tot}=1$. In order to get 
also excited (singlet) states within $S^z_{tot}=0$ sector we evaluate the g.s. 
eigen-function $|\psi_0\rangle$ and then construct effective Hamiltonian for the 
excited states $H_1=H-E_0|\psi_0\rangle\langle\psi_0|$ 
\cite{wang18}, and then repeat the standard DMRG algorithm for 
$H_1$. The requirement of orthogonality is, however, difficult to meet for 
excited states which are (due to o.b.c.) edge states, e.g., as within 1D 
dimerized regime.

%================================================================================
\section{One-dimensional Heisenberg model}

%--------------------------------------------------------------------------------
\begin{figure}[!ht]
\centering
\includegraphics[width=0.9\columnwidth]{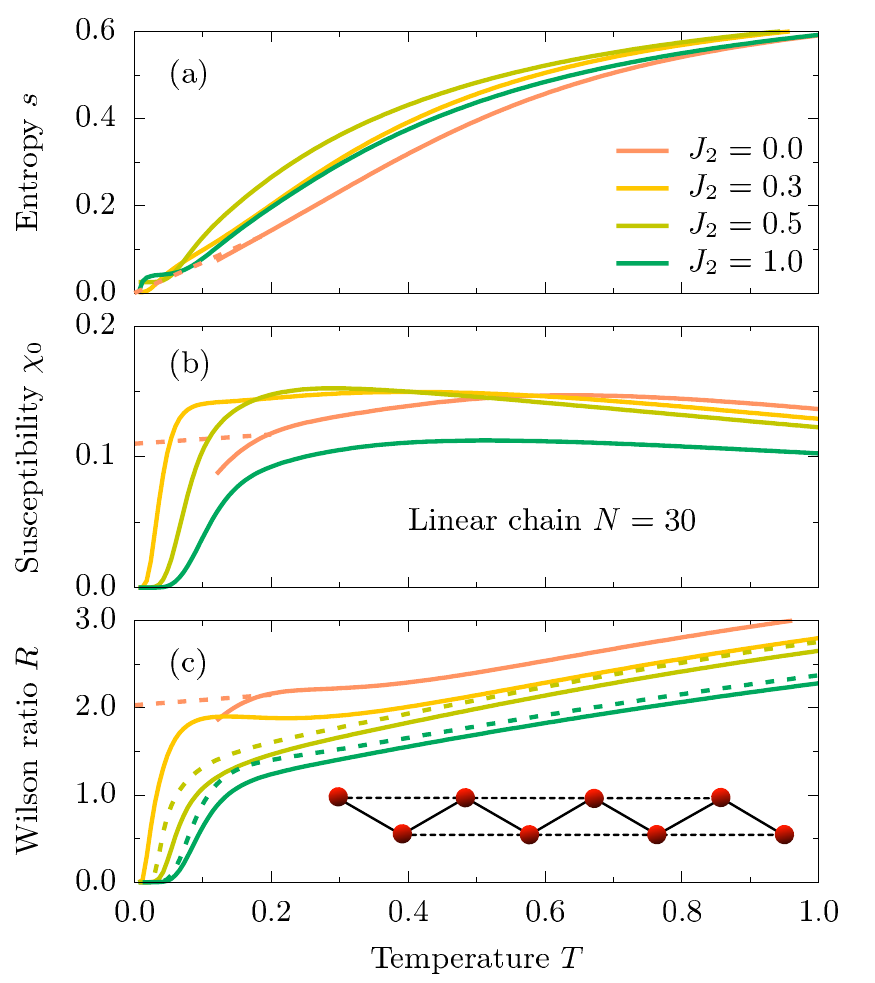}
\caption{ Results in the $J_1$-$J_2$ Heisenberg chain for: (a) entropy $s(T) $, 
(b) susceptibility $\chi_0(T)$, and (c) Wilson ratio $R(T)$, as obtained via 
FTLM on $N=30$ sites for different $J_2 = 0.0 - 1.0$. The dashed lines at 
$J_2=0$ represent the extension to $N \to \infty$, while for $J_2=0.2, 0.3$ they 
denote modified $R(T)$ evaluated with reduced $\tilde s(T)$. The inset in (c) 
represents a sketch of $J_1$ (solid line) and $J_2$ (dashed line) in 1D 
Heisenberg chain.}
\label{llj12}
\end{figure}
%--------------------------------------------------------------------------------

We consider first the 1D $J_1$-$J_2$ HM, which can serve as the reference for 
further discussion of 2D HM results. The AFM isotropic $S=1/2$ $J_1$-$J_2$ HM is 
given by
\begin{equation}
H= \sum_{i} \left[ J_1 {\bf S}_i \cdot {\bf S}_{i+1} + J_2 {\bf S}_i \cdot {\bf S}_{i+2} \right]\,,
\label{eqllj12}
\end{equation}
where we further on put $J_1=J=1$ as the unit of energy. We will investigate 
with FTLM only $J_ 2\geq 0$ case on systems of finite length $N$ with p.b.c. 
Thermodynamic properties are well known and understood for simple $J_2=0$ 
Heisenberg chain \cite{johnston00}, as well the g.s. and the triplet excited 
state for the frustrated chain with $J_2>0$ \cite{white96}. Beyond critical 
$J_2 > J_2^* \sim 0.241$ the g.s. is dimerized ($N_d =2$) in the thermodynamic 
limit \cite{white96}. At the same time, lowest excited states are degenerate 
triplets and singlets with the gap $ \Delta_ t = \Delta_s$, consistent with the 
unbound spinons as elementary excitations. 

Numerical results for $s(T)$, $\chi_0(T)$ and finally $R(T)$, as obtained on a 
system with $N=30$ sites, are presented in Fig.~\ref{llj12} for different $0 \leq J_2 \leq 1$:

\noindent a) For the simple $J_2=0$ chain we get $s(T) \sim \gamma T$ in very 
broad range $T < 0.6$. Finite-size effects are most pronounced in this case, so 
that below $T < T_{fs} \sim 0.2$ we get $s < 0.1$ and finite-size effects 
prevent any further firm conclusions. Still, for $T>T_{fs}$ numerical results 
are consistent with analytical and previous numerical results, 
in particular with the known limit $R_0 = 2$  \cite{johnston00}. Moreover, it is remarkable that 
$R(T)$ is nearly constant in a wide range $T < 0.6$. 

\noindent b) The gap becomes pronounced for the Mazumdar-Ghosh point $J_2=0.5$ 
and even more for $J_2 = 1.0$ (where $\Delta_t \sim 0.25$ \cite{white96}). In 
the gapped case FTLM finite-size effects are less pronounced, and one can
expect $T_{fs} \to 0$. In fact, for the $J_2=0.5$ and $J_2=1.0$ results appear 
size-independent for reached $N=30$, apart from the dimerization degeneracy 
$N_d=2$ leading via Eq.~(\ref{s0}) to $s_0>0$. The latter has influence on the 
$R(T \sim 0)$, so we present in Fig.~\ref{llj12} also the result taking into 
account subtracted $\tilde s(T)$. In both analyses the behavior is consistent 
with $R_0=0$. For the $J_2=0.5$ and $J_2=1.0$ modified results are still 
consistent with vanishing $R(T< \Delta) \propto T^\eta$ with $\eta\geq 1$, but 
this behavior remains to be clarified. For the marginal case 
$J_2 = 0.3 \sim J_2^*$, the behavior of all quantities is similar to $J_2=0$, 
except that we find larger $\gamma$ and consequently also smaller $T_{fs}$ .

\begin{figure}[!t]
\centering
\includegraphics[width=0.9\columnwidth]{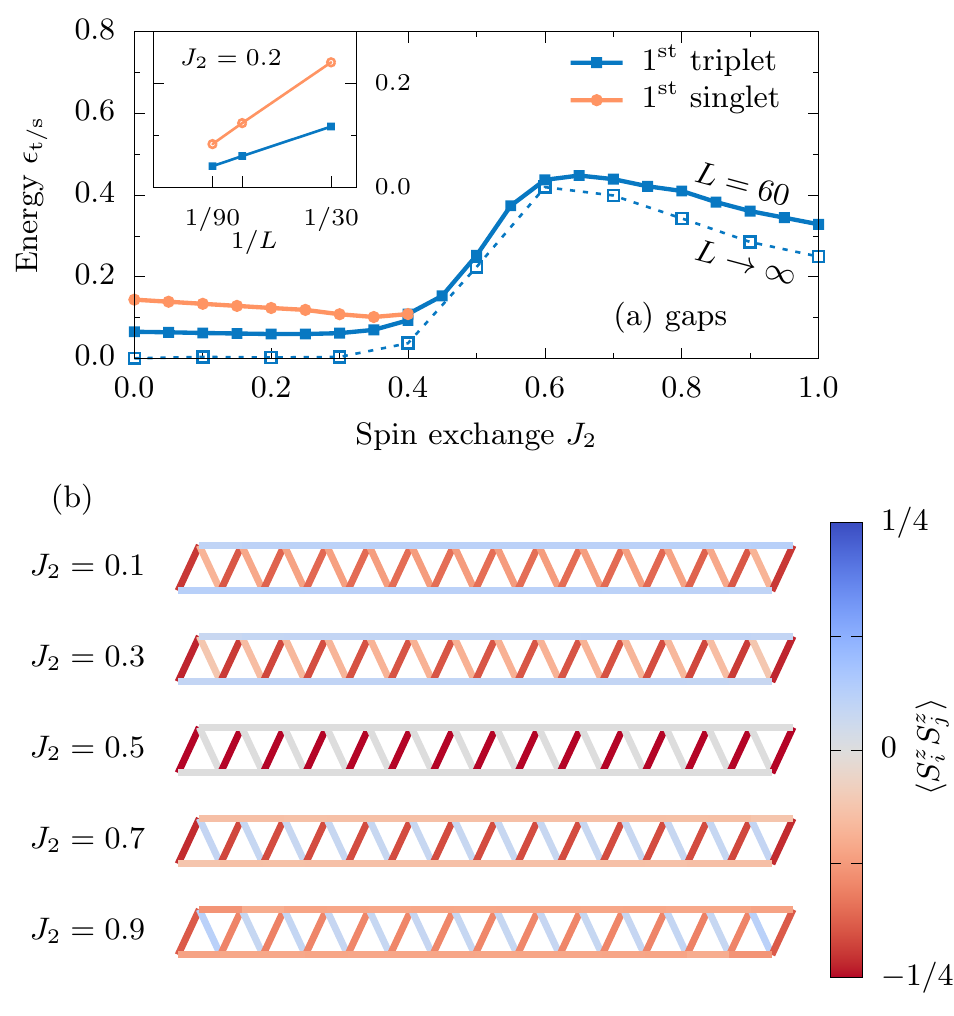}
\caption{(a) Lowest triplet $\epsilon_t$ and singlet $\epsilon_s$ excitations 
vs. $J_2$, as obtained via DMRG in the chain of $N=60$ sites, with the inset 
showing the scaling of $\epsilon_{t/s}$ vs. $1/N$ for $J_2=0.2$. (b) 
Corresponding g.s. spin correlations $\langle S^z_i S^z_j \rangle$ on particular 
bonds.}
\label{chj}
\end{figure}

It is instructive to investigate in connection with finite-size effects also 
lowest triplet and singlet excitations in the model. While triplet excitations 
have been in detail studied using DMRG already in Ref.~\cite{white96}, to 
establish singlet excitations requires more care, see Sec.~III. In Fig.~\ref{chj}
(a) we present the DMRG (with o.b.c.) $N=60$ result for excitations: lowest 
triplet $\epsilon_t$ and lowest singlet $\epsilon_s$ vs. $J_2$, together (as the 
inset) with their $1/N$ scaling in the gapless regime $J_2 = 0.2 <J_2^*$. Due to 
o.b.c. DMRG is unable to properly resolve the dimerized partner of g.s. since it 
represents in open chain excited edge states. Hence, we present in 
Fig.~\ref{chj}(a) the first singlet excited state only for $J_2 \leq 0.4$. 
Still, DMRG results confirm that no other singlet is stable below the triplet
for $J_2 > J_2^*$, unlike seen later on in 2D SL models.

%--------------------------------------------------------------------------------
%--------------------------------------------------------------------------------

In Fig.~\ref{chj}(b) we display also DMRG results for g.s. bond spin 
correlations $\langle S^z_i S^z_j \rangle$. It is also apparent that for 
$J_2>J_2^*$ the g.s. is dimerized (in n.n. bond correlations), whereby the 
particular case is $J_2 =0.5$ with alternating n.n. correlations 
$\langle S^z_i S^z_j \rangle =-1/4$ and $0$. Stronger correlations remain AFM in 
the whole $J_2>J_2^*$, while it is easy to recognize the change of character 
of weaker bonds from AFM correlations for $J_c^* <J_2 < 0.5$, to erromagnetic  ones 
for $J_2>0.5$. 

%================================================================================
\section{Planar frustrated Heisenberg models}

%================================================================================
\subsection{$J_1$-$J_2$ Heisenberg model on kagome lattice}

\begin{figure}[!ht]
\centering
\includegraphics[width=0.9\columnwidth]{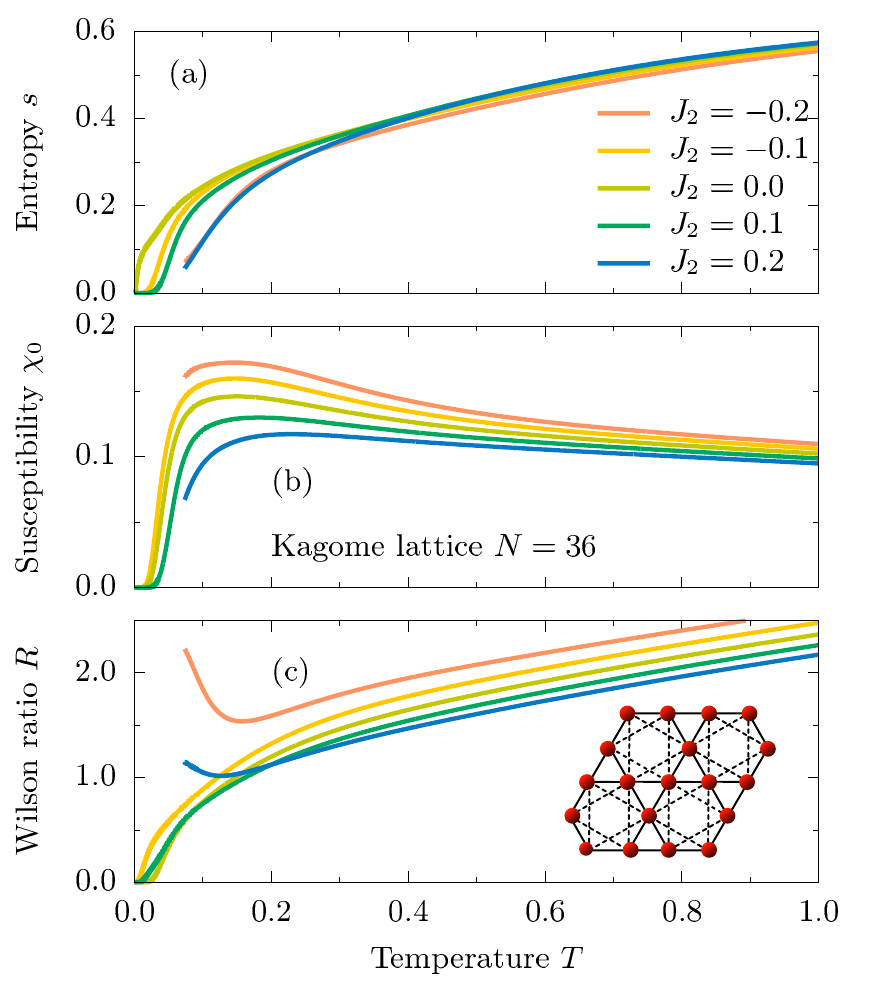}
\caption{$s(T) $, $\chi_0(T)$ and $R(T)$ within the $J_1$-$J_2$ HM on KL, 
obtained via FTLM on $N=36$ sites, for different $|J_2| \leq 0.2$. The inset in 
(c) represents a sketch of the $J_1$ (solid line) and $J_2$ (broken line) 
connections in KL.}
\label{klj12}
\end{figure}

HM on KL is the prototype model for the existence of SL in planar models. It has 
been the subject of numerous studies, devoted mostly to the g.s. using ED 
\cite{mila98,budnik04,lauchli11,lauchli19}, series expansion \cite{singh07}, 
DMRG \cite{yan08,depenbrock12,liao17} and variational methods 
\cite{iqbal11,iqbal13}. We consider here the extended model with p.b.c., 
involving also the n.n.n. exchange $J_2$ as shown in the inset of 
Fig.~\ref{klj12}(c),
\begin{equation}
H= J_1 \sum_{\langle ij\rangle } {\bf S}_i \cdot {\bf S}_j + 
J_2 \sum_{\langle \langle il \rangle \rangle} {\bf S}_i \cdot {\bf S}_l\,,
\label{2dj12}
\end{equation}
whereby the role of $J_2>0$, as well as $J_2<0$, is to reestablish the magnetic LRO 
\cite{kolley15}. The basic HM on KL has been the clearest case for a dominant 
role of low-lying singlet excitations over the triplet ones 
\cite{singh07,singh08,lauchli19}. The latter fact and related large entropy, 
persistent at low $T \ll 1$, has been well captured within block-spin 
\cite{subrahmanyam95,mila98,budnik04} and recently within related reduced-basis 
approach \cite{prelovsek19}, whereby singlet excitations can be attributed 
to chiral fluctuations, distinct from (higher-energy) triplet excitations.

%--------------------------------------------------------------------------------
%--------------------------------------------------------------------------------

Thermodynamic quantities for the basic $J_2=0$ HM on KL have been calculated via 
FTLM previously \cite{schnack18} up to the size $N=42$. Here we extend the 
study, evaluating via FTLM also for $J_2 \neq 0$ for $N = 36$. Results in 
Fig.~\ref{klj12} reveal that increasing $|J_2| > 0$ suppresses strongly $s(T \ll J_1)$ 
while leaving $\chi_0(T)$ less affected (at least for $T>T_{fs}$). Results 
for $J_2 =\pm 0.2$ indicate on divergent $R_0 \to \infty$ consistent with the 
emergent magnetic LRO \cite{kolley15,prelovsek19}. On the other hand, at 
$|J_2| \leq 0.1$ the behavior of $\chi_0(T)$ and $s(T)$ are consistent with 
finite triplet gap $\Delta_t \sim 0.15$ and smaller or even vanishing singlet 
gap $\Delta_s < \Delta_t$.

%--------------------------------------------------------------------------------
\begin{figure}[!ht]
\centering
\includegraphics[width=0.9\columnwidth]{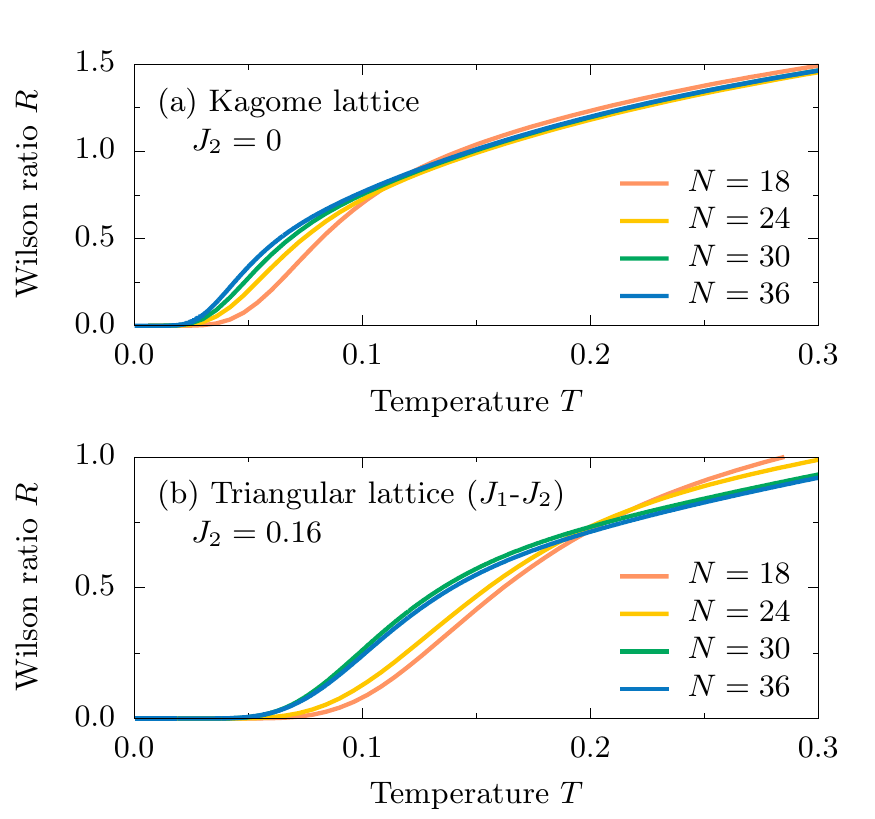}
\caption{Finite-size comparison of $R(T)$ 
within the SL regimes for:  (a) basic KL model with $J_2=0$, and (b)  $J_1$-$J_2$ model on TL.
Results are obtained via FTLM on lattices with $N=18 - 36$ sites. }
\label{KL_size}
\end{figure}
%--------------------------------------------------------------------------------

Results in Fig.~\ref{klj12} are quite robust against finite-size
effects, in particular in the presumable SL regime. To substantiate this we show in Fig.~\ref{KL_size}(a)
$R(T)$ in a low-$T$ regime for the KL at $J_2 =0$, as obtained  via FTLM
on lattices of quite different sizes $N = 18 - 36$. It should be stressed that we do not pretend to 
perform a proper finite-size scaling, since considered lattices are not just of different sizes, but also
of different shapes (due to requirement of p.b.c.), e.g., lacking some (rotational) symmetries etc. 
Still, results for $R(T)$ in Fig.~\ref{KL_size}(a) (as well for the TL on Fig.~\ref{KL_size}(b),
discussed further on) reveal quite systematic evolution of $R(T)$ consistent with vanishing
$R(T \to 0)$.

%--------------------------------------------------------------------------------
\begin{figure}[!ht]
\centering
\includegraphics[width=0.95\columnwidth]{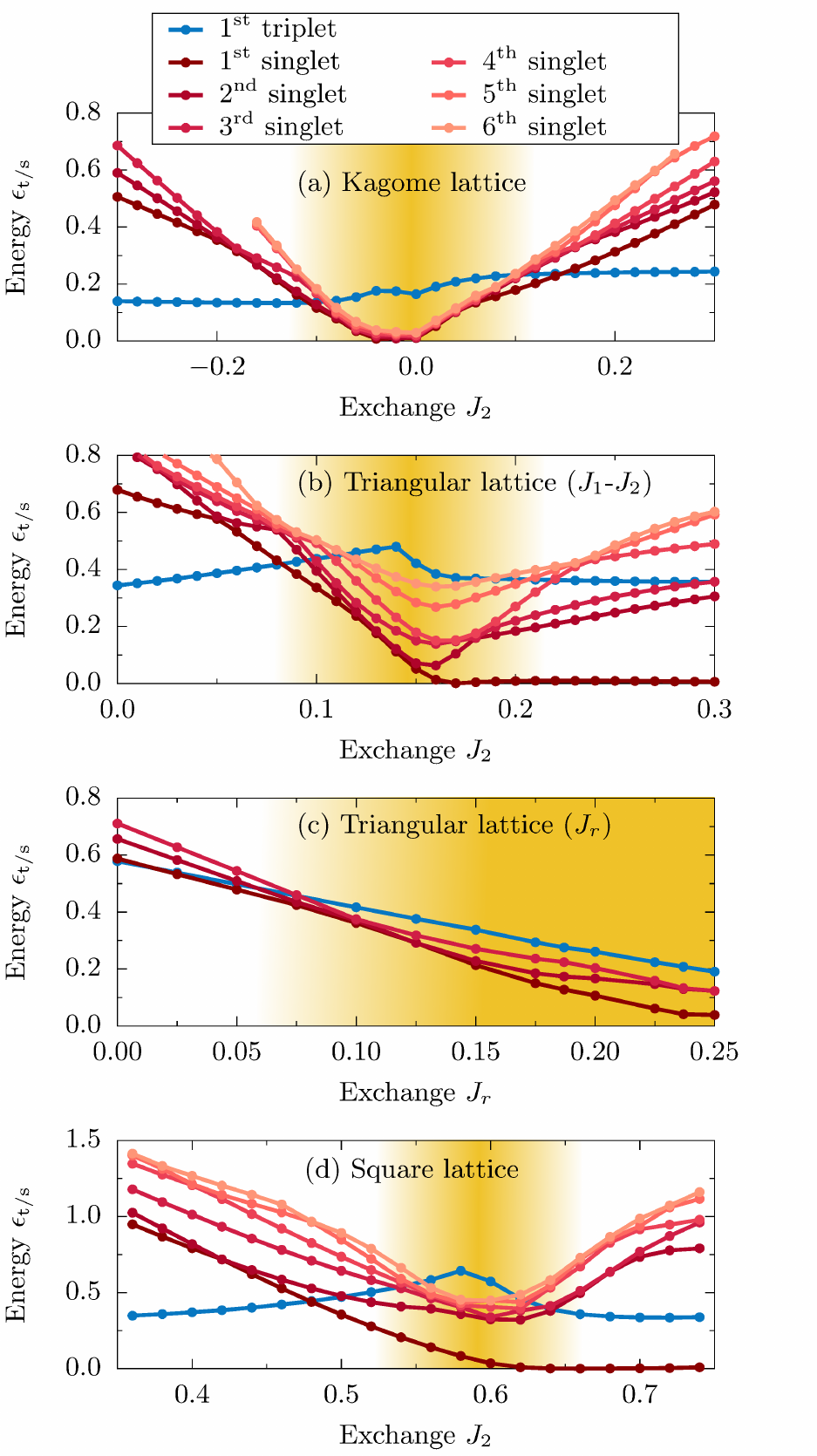}
\caption{Lowest triplet excitation $\epsilon_t$ and (nondegenerate) singlet 
excitations $\epsilon_{s,i}$ ($i=1,2\cdots,6$) vs. $J_2$ for different planar 
$J_1$-$J_2$ HM models on: (a) KL , (b) TL, and (d) SQL, as obtained with ED on 
$N=36$ sites, and (c) $\epsilon_t$ and $\epsilon_{s,i=1,2,3}$ vs. $J_r$ on TL 
with ring exchange, obtained on $N=28$ sites.}
\label{levels}
\end{figure}
%--------------------------------------------------------------------------------

The transition from the singlet-dominated SL regime to the phases with magnetic 
LRO one can monitor also via low-lying levels in considered systems. In 
Fig.~\ref{levels} we present the evolution of excitation energies for lowest 
lying triplet $\epsilon_t$ as well as several low-lying excited singlets 
$\epsilon_{s,i}$ ($i=1,2,\cdots, 6$), as obtained via ED on $N=36$ sites, and in 
part for $N=32$ sites for the HM on TL with ring exchange. It should be pointed 
out that we monitor only nondegenerate excited states, whereby in general the 
degeneracy is present and depends on particular lattice and related p.b.c. The 
level evolution, plotted vs. $J_2$ (or $J_r$ discussed later on) serves primarily 
as another test where to expect SL with macroscopic number (in the limit 
$N \to \infty$) of singlet excitation below the triplet ones, but also to locate 
transitions between different regimes.

In Fig.~\ref{levels}(a) the level scheme for KL is consistent with the previous ED 
studies of ($J_2=0$) KL model \cite{singh07,singh08,lauchli19} which reveal a 
massive density of singlet levels with $\epsilon_s \sim 0$ below the lowest 
triplet one $\epsilon_t$. Introducing $|J_2| > 0$ reduces the degeneracy and 
might lead to $\Delta_s > 0$ even in the $N \to \infty$ limit. Still, a large 
density of singlet levels appear below the triplet in a wide (SL) range 
$J_2^{c1} < J_2 < J_2^{c2} $ where $J_2^{c1} \sim -0.1, J_c^{c2} \sim 0.1$ from 
Fig.~\ref{levels}(a) and we define $J_2^{c1,c2}$ with the crossing of (all) 
lowest $\epsilon_{s,1-6} < \epsilon_t$. We note that marginal $J_2^{c1,c2}$ are 
consistent with Fig.~\ref{klj12} where $J_2 = \pm 0.2$ already reveal
magnetic LRO with $R_0 \to \infty$.

%================================================================================
\subsection{$J_1$-$J_2$ Heisenberg model on triangular lattice}

While numerical studies for the basic ($J_2=0$) HM on TL 
\cite{bernu94,capriotti99,white07} confirm magnetic LRO with moments pointing into 
$120^0$-angle directions, modest additional frustration with $J_2 > 0$ allows 
for the possibility of SL g.s., with the evidence for either gapless 
\cite{kaneko14} or gapped SL \cite{zhu15,hu15,iqbal16,wietek17} in the 
intermediate regime $J_2 \sim 0.15$. Beyond that, for $J_2 > 0.2$
stripe AFM is expected. Thermodynamic (and some dynamic) quantities for the 
$J_1$-$J_2$ HM, Eq.~(\ref{2dj12}), on TL have been recently calculated using 
FTLM \cite{prelovsek18} up to $N=30$ sites and employing the reduced-basis 
approach \cite{prelovsek19}, whereby the similarity of $s(T)$, $\chi_0(T)$ and 
$R(T)$ with the basic HM on KL in the SL regime in both models has been attributed
to chiral fluctuations dominating low-$T$ excitations. 

%--------------------------------------------------------------------------------
\begin{figure}[!ht]
\centering
\includegraphics[width=0.9\columnwidth]{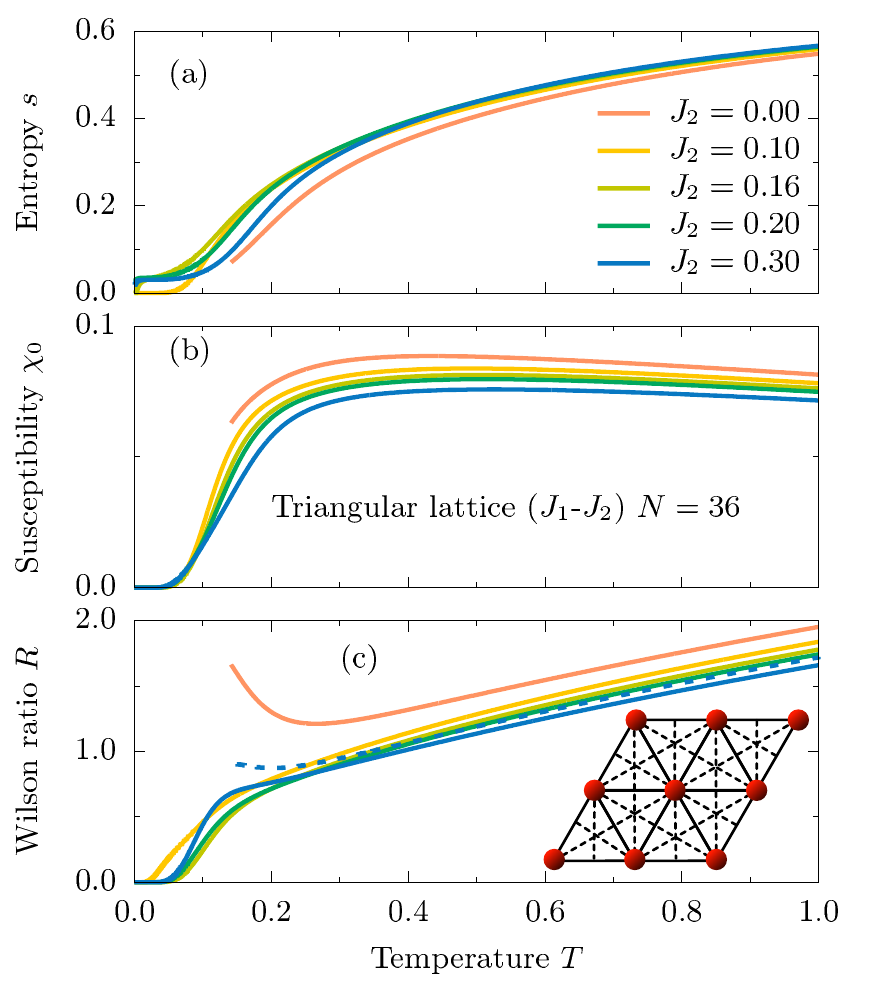}
\caption{ $s(T)$, $\chi_0(T)$ and $R(T)$ within $J_1$-$J_2$ HM on TL, 
obtained  via FTLM on $N=36$ sites for different $J_2 \leq 0.3$. The dashed line for 
$J_2=0.3$ represents result using reduced $\tilde s(T)$. The inset in (c) 
represents a sketch of the $J_1$ (solid line) and $J_2$ (broken line) 
connections in TL.}
\label{tlj12}
\end{figure}
%--------------------------------------------------------------------------------

Here we upgrade previous FTLM studies with the calculation of $J_1$-$J_2$ HM on 
TL with $N=36$ sites. Results in Fig.~\ref{tlj12} are qualitatively 
consistent with previous ones for $N=30$ \cite{prelovsek18} but due to larger 
size and consequently smaller $T_{fs}$ results are reliable to smaller entropy 
$s(T)$ and more evidently reveal diverging $R(T)$ below $T\sim 0.2$ for $J_2 \sim 0 $, where 
the g.s. possesses magnetic LRO. A similar behavior is expected for $J_2 > 0.2$ 
where the stripe AFM g.s. has been established \cite{kaneko14}. In reachable 
system $N=36$ the upturn of $R(T)$ is partly masked by finite-size $s_0 > 0$, 
Eq.~(\ref{s0}), due to the degeneracy $N_d>1$ of striped magnetic LRO, evident in 
Fig.~\ref{tlj12}(a) at $J_2=0.2$ and $0.3$. Taking into account in 
Eq.~(\ref{rw}) the reduced $\tilde s = s-s_0$, we obtain for $J_2=0.3$ again the 
indication for the upturn of $R(T)$ consistent with g.s. magnetic  LRO. Still, in the most 
important intermediate regime $0.1 < J_2 < 0.2$ the increase of $s_0$ and at the 
same time fast decrease of $\chi_0(T \to 0)$ (indicating a finite triplet gap 
$\Delta_t>0$) leads to vanishing $R_0 =0$ and is consistent with interpretation
with the SL g.s. \cite{kaneko14,zhu15,hu15,iqbal16,wietek17}.  

Again, in this intermediate regime 
results are most robust against finite-size effects. In Fig.~\ref{KL_size} we
display the results for $R(T)$ for particular $J_2 =0.16$, as obtained on quite
different sizes $N = 18 - 36$, whereby all presented lattices are not optimal with 
respect to lattice symmetries. Nevertheless, low-$T$ variation of $R(T)$ appears
at least qualitatively consistent for all $N$. 

In Fig.~\ref{levels}(b) we plot the corresponding evolution of excitations vs. 
$J_2$, as obtained with ED on $N=36$ lattice. The triplet gap apparently remains 
substantial, i.e. $\epsilon_t >0.38$ for considered $N$ in the whole range of 
$J_2< 0.3$. Still, singlet excitations $\epsilon_{s,1-6}$ all cross 
$\epsilon_{t}$ for small $J_2 \sim 0.1$. This leads effectively to g.s. level 
crossing $\epsilon_{s,1}=0$ at $J_2 \sim 0.17$ exchanging the character of the 
g.s. into a striped AFM. But most important, in the intermediate range 
$0.1 < J_2 < 0.17$, which should be the relevant SL regime, singlet-excitation
collapse is consistent with the conclusions from thermodynamics in 
Fig.~\ref{tlj12} and $R_0=0$. It should be, however, acknowledged that the 
singlet collapse is not as pronounced as for basic ($J_2 \sim 0$) HM on KL in 
Fig.~\ref{levels}(a).

%================================================================================
\subsection{Heisenberg model with ring exchange on triangular lattice}

\begin{figure}[!ht]
\centering
\includegraphics[width=0.9\columnwidth]{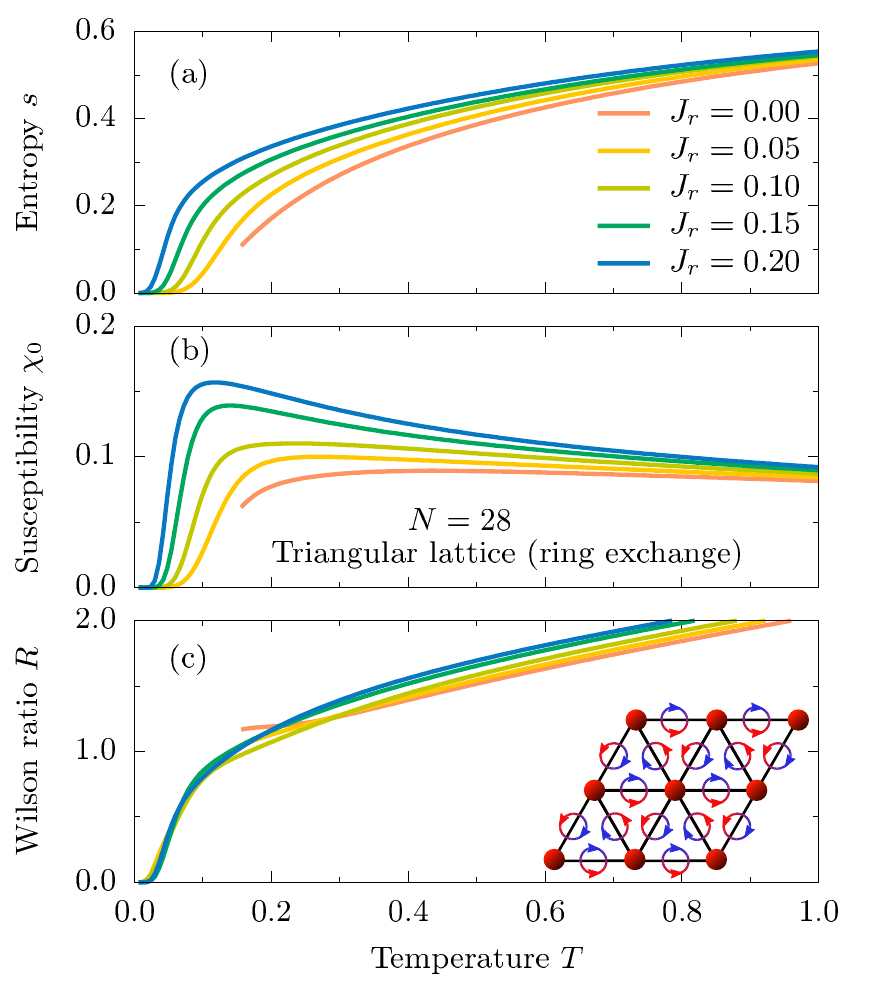}
\caption{ $s(T)$, $\chi_0(T)$ and $R(T)$ within HM on TL, including ring 
exchange, obtained via FTLM on $N=28$ sites for different $0 \leq J_r \leq 0.2$. 
The inset in (c) represents a sketch of the $J$ (solid line) connections and 
the ring exchange $J_r$ (circle) in TL.}
\label{tljr}
\end{figure}

While $J_1$-$J_2$ HM on TL is conceptually simple, it is less obvious to justify 
in connection with experiments and with more basic models. The organic SL 
materials \cite{shimizu03,shimizu06,itou10,zhou17} and 1T-TaS$_2$ 
\cite{klanjsek17,kratochvilova17,law17,he18} are closer to the metal-insulator 
transition where simple $S=1/2$ n.n. HM is presumably not enough. Assuming as 
the starting point the single-band Hubbard model on the insulator side of the 
Mott transition $U > U_c$ the lowest correction to the n.n. HM comes in the form 
of the ring exchange term \cite{misguich99,motrunich05,yang10,nakamura14},
\begin{eqnarray}
H &=& J \sum_{\langle ij\rangle } {\bf S}_i \cdot {\bf S}_j + H_r\,,
\end{eqnarray}
with
\begin{eqnarray}
H_r &=& \frac{J_r}{2} \sum_{\langle ijkl \rangle} (P_{ijkl} + P_{lkji} ) \sim
J_r \sum_{\langle ijkl \rangle} \bigl [ ({\bf S}_i \cdot {\bf S}_j) ({\bf S}_k \cdot {\bf S}_l) \nonumber \\
&+& ({\bf S}_i \cdot {\bf S}_l) ({\bf S}_j \cdot {\bf S}_k) 
- ({\bf S}_i \cdot {\bf S}_k) ({\bf S}_j \cdot {\bf S}_l) \bigr]\,,
\label{hring}
\end{eqnarray}
where $\langle ijkl \rangle$ are taken over different four-cycles on TL, as 
shown in the inset of Fig.~\ref{tljr}(c). $H_r$, Eq.~(\ref{hring}), has been 
confirmed as the leading correction in the numerical study of the half-filled 
Hubbard model \cite{yang10} in the insulating regime where 
$J_r \sim 80 t^4/U^3 \sim (20 t^2/U^2) J < 0.2 J$ \cite{nakamura14}, taking into 
account that the Mott insulator on TL requires $U > U_c \sim 8t-10t$ and 
$J \sim 4t^2/U$. It should be also mentioned that in Eq.~(\ref{hring}) in our numerical study 
we neglect the (small) corrections to the n.n. exchange term and corresponding $J$,
which emerge in the same order of the $t/U$ expansion. 

%--------------------------------------------------------------------------------
%--------------------------------------------------------------------------------

It has been already proposed that modest ring exchange $J_r > 0$ on TL destroys 
the magnetic LRO and induces SL g.s. \cite{misguich99,motrunich05}, including the 
observation of several possible singlet excitations below the lowest triplet 
one. In Fig.~\ref{tljr} we present results for the HM on TL with $J_r > 0$, 
Eq.~(\ref{hring}), as obtained via FTLM on $N=28$ sites (smaller size due to 
more complex $H$). It is evident that $J_r > 0$ steadily increases low-$T$ 
entropy $s(T)$, while increasing $\chi_0(T)$. Resulting $R(T)$ looses magnetic 
LRO  character already for $J_r \geq 0.05$ being followed by SL-like regime with 
vanishing $R_0 \to 0$. 

The same message follows from the consideration of lowest levels on
$N=28$ lattice, presented in Fig.~\ref{levels}(c). Analogous to 
Figs.~\ref{levels}(a) and \ref{levels}(b), there is a clear collapse of singlet 
levels $\epsilon_{s,1-3}$ (here we employ a ${\bf q}$-resolved code and cannot 
monitor all singlet excitations) below the triplet one $\epsilon_t$ for 
$J_r > 0.1$. In the latter regime $\epsilon_t $ represents already a reasonable 
estimate of the limiting $N \to \infty$ triplet gap $\Delta_t >0$ 
\cite{misguich99}, whereas to establish a proper singlet gap (the lowest singlet 
in $N \to \infty$ limit) $\Delta_s < \Delta_t$ requires more detailed 
finite-size analysis. 

%================================================================================
\subsection{$J_1$-$J_2$ Heisenberg model on square lattice}

%--------------------------------------------------------------------------------
\begin{figure}[!ht]
\centering
\includegraphics[width=0.9\columnwidth]{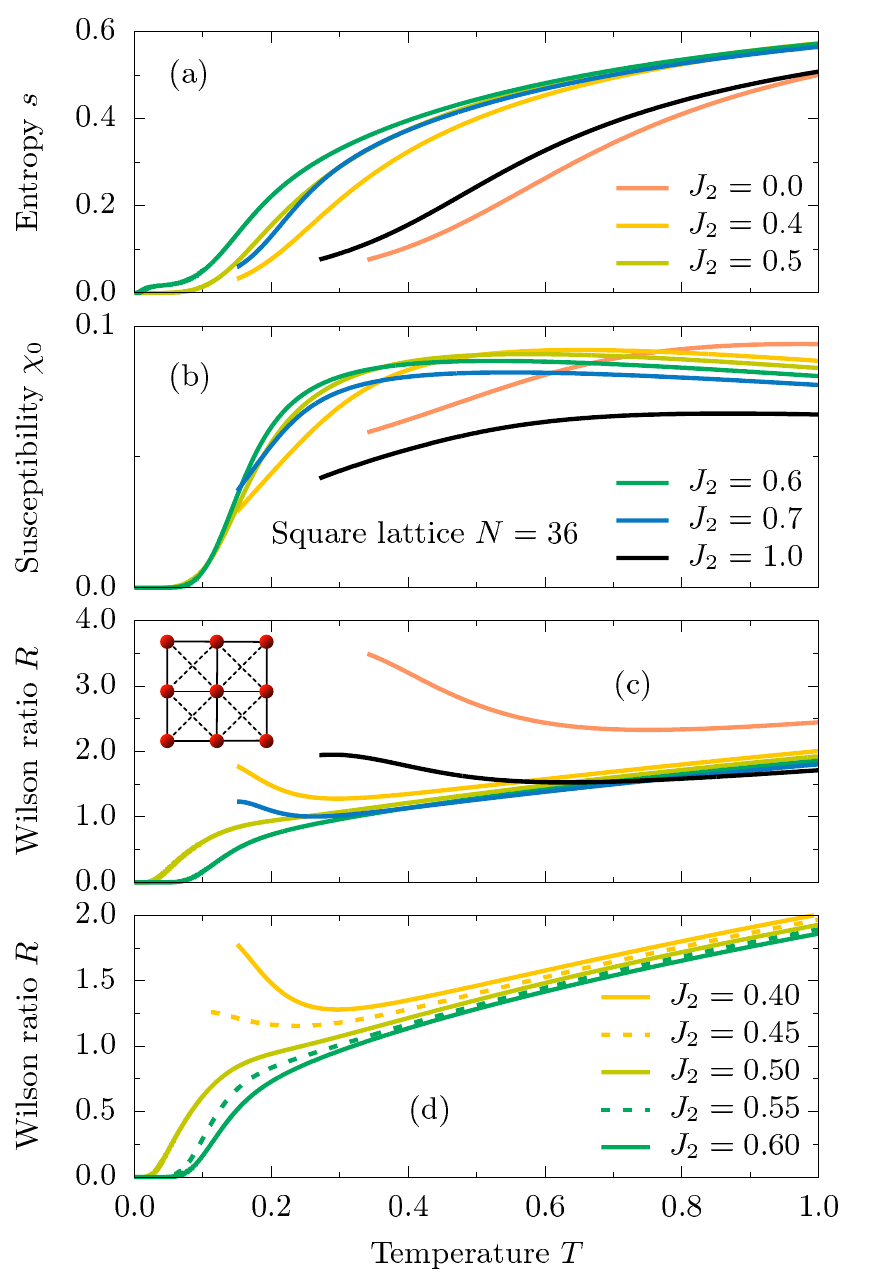}
\caption{ (a) $s(T)$, (b) $\chi_0(T)$ and (c) $R(T)$ within $J_1$-$J_2$ HM on SQL, 
obtained  via FTLM on $N=36$ sites for different $0 \leq J_2 \leq 1.0$. In (d)
$R(T)$ results are presented within the expanded intermediate regime
$0.4 \leq J_2 \leq 0.6$. }
\label{sqj12}
\end{figure}
%--------------------------------------------------------------------------------

Finally, we turn to the $J_1$-$J_2$ HM, Eq.~(\ref{2dj12}) on SQL. The latter has 
been one of first considered for the possible (plaquette) VB solid g.s. at 
intermediate $J_2 \sim 0.5$ \cite{capriotti00,mambrini06,doretta14,morita16,zhao19}, but 
also for the SL g.s. \cite{jiang12,gong14,morita15,wang18,liu18}. Results for 
corresponding thermodynamic quantities presented in Fig.~\ref{sqj12}c,d are 
consistent with the diverging $R_0 \to \infty$ indicating magnetic LRO outside quite 
narrow parameter regime, i.e. outside $0.5 \leq J_2 \leq 0.6$. In the latter 
regime we again find substantial entropy $s(T \ll 1)$ and consequently 
$R_0 \to 0$, whereby for $J_2 \sim 0.6$ there are already some indications for 
possible degeneracy $s_0>0$ which could be in favor of broken translational 
symmetry, e.g., a plaquette VB solid \cite{capriotti00,mambrini06,
doretta14,morita16,zhao19}.

Caveats for the SL interpretation emerge also when considering the excitation 
evolution vs. $J_2$ [see Fig.~\ref{levels}(d)], as obtained from ED results on 
$N=36$ cluster. For given system size, the singlet levels reveal 
$\epsilon_{s,1-6} < \epsilon_t$ only in a very narrow regime $0.55< J_2 < 0.62$. 
Even then, higher singlets (apart from $\epsilon_{s,1}$) are not well below 
$\epsilon_t$. Consistent with previous works 
\cite{capriotti00,mambrini06,jiang12,gong14,morita15,morita16,wang18,liu18} 
level scheme indicates on a change of the g.s. character for $J_2 > 0.6$. As a 
consequence, the SL in the intermediate regime, and even more on the 
singlet-dominated regime is less conclusive, and other options 
\cite{wang18,zhao19} have to be also considered. 

%================================================================================
\section{Conclusions}

Thermodynamic quantities: entropy density $s(T)$ (together with directly related 
specific heat $C_V(T) = T ds/dT$, not presented in this paper), uniform 
susceptibility $\chi_0(T)$, and consequently $T$-dependent Wilson ratio $R(T)$, 
offer another view on properties of frustrated spin models. We considered here 
prototype 2D isotropic $S=1/2$ HM, which are at least in some parameter regimes 
best candidates for the SL g.s. For comparison, we investigated in the same 
manner also simplest 1D HM which can serve as reference for some concepts and 
scenarios. 

$R(T)$, in particular its low-$T$ variation, is the quantity which 
differentiates between different scenarios. Whereas 2D systems with magnetic LRO 
can be monitored via $R_0 \to \infty$, we are more interested in the SL regimes 
with g.s. without magnetic LRO and even without any broken translational 
symmetry which could be classified as VB solid. As prototype case we 
present results for 1D $J_1$-$J_2$ HM which does not have magnetic LRO, but offers 
already two firm scenarios: a) the gapless regime for $J_2< J_2^*$ with spinons 
(or 1D SFS) as elementary excitations, and consequently finite 
$R_0 = R(T \to 0) \sim 2$ (for $J_2 \sim 0$), b) a gapped regime for $J_2>J_2^*$ 
with dimerized g.s. (being the simplest 1D form of VB solid) apparently also 
with $R_0 =0$, although not yet fully resolved variation $R(T \to 0)$.

SL regimes in considered 2D frustrated isotropic $S=1/2$ HM are in our study 
located via enhanced low-$T$ entropy $s(T)$ and gapped (or at least reduced) 
$\chi_0(T)$, resulting in vanishing $R_0=0$. Similar information and criterion 
(although less well defined) emerges from the excitation spectra, when 
differentiating singlet and triplet (or even higher $S_{tot}>1$) excitations 
over the $S_{tot}=0$ g.s. Most evident cases for such VB (dimer) liquid scenario 
appears within the KL around $J_2 \sim 0$. Analogous, although somewhat less 
pronounced, case is obtained within HM on TL with ring exchange $J_r > 0.1$ and 
for the $J_1$-$J_2$ HM on TL in the intermediate regime $0.1<J_2 < 0.17$. For 
such systems the level evolution as well as $R(T)$ reveal massive density of 
singlet states below the lowest triplet excitation. On the other hand, the 
situation in the HM on SQL in the narrow regime $J_2 \sim 0.6$ is less clear-cut 
in this respect, since singlets are not well below the lowest triplet. 

Vanishing $R_0$ does not support the scenario of SL with large (or even 
Dirac-cone) spinon Fermi surface, which would require finite $R_0 >0$ (as in 1D 
HM), although our finite-size studies should be interpreted with care and cannot 
give a final answer to this problem. Still, emergent scenario of VB
liquid should be critically faced with the possibility of VB solid. In the latter case the 
g.s. should be (due to broken  translational symmetry) degenerate with finite $N_d > 1$ 
(in the thermodynamic  limit $N \to \infty$). We find clear numerical evidence for $N_d > 1$ within the 
$J_1$-$J_2$ HM on TL for $J_2 > 0.2$, but in this case it is consistent 
with the striped magnetic LRO. Some indication for $N_d > 1$ appears also for the 
SQL at $J_2 \sim 0.6$, which might support the existence of (plaquette) VB 
solid \cite{gong14,wang18} in this regime instead of SL (without broken 
translational symmetry).  Results in the (presumable) SL regimes also indicate 
on finite triplet gaps $\Delta_t >0$ while singlet gaps are either finite
$\Delta_s > 0$ or vanishing $\Delta_s \sim 0$ (for the $J_2 \sim 0$ KL model), 
but evidently  $\Delta_s < \Delta_t$. To establish (or exclude) possible $N_d>1$ and to 
determine $\Delta_s>0$ beyond doubt still requires further studies.

Finally, it should be stressed that evaluated thermodynamic quantities are (at 
least in principle) measurable in experimental realizations of SL materials. 
$s(T)$ is accessible via measured magnetic specific heat $C_V(T)$ and uniform 
susceptibility $\chi_0(T)$ via macroscopic d.c. or/and Knight-shift measurement. 
Since known SL materials are characterized by modest exchange $J$, properties 
can be measured in the wide range $T \lesssim J$. This offers the 
possibility of critical comparison with model results, whereby considered 
isotropic HM might still miss some ingredients relevant for the low-$T$ 
behavior, in particular the Dzyaloshiniskii-Moriya interaction, the disorder 
influence, and the inter-layer coupling. 
 
\acknowledgments P.P. is supported by the program P1-0044 and project N1-0088 of 
the Slovenian Research Agency. K. M. and T. T. are supported by MEXT, Japan, as 
a social and scientific priority issue (creation of new functional devices and 
high-performance materials to support next-generation industries) to be tackled 
by using a post-K computer. T.T. is also supported by the JSPS KAKENHI 
(No. JP19H05825). The numerical calculation was partly carried out at the 
facilities of the Supercomputer Center, the Institute for Solid State Physics, 
the University of Tokyo, at the Yukawa Institute Computer Facility, Kyoto 
University, and at the Wroclaw Centre for Networking and Supercomputing. J. H. 
acknowledges grant support by the Polish National Agency of 
Academic Exchange (NAWA) under contract PPN/PPO/2018/1/00035.

%================================================================================
\bibliography{manuwilson}%.bbl}
\end{document}